\documentclass[]{article}

\renewcommand{\arraystretch}{1.3}

\catcode`\@=11
\def\marginnote#1{}

\newcount\hour
\newcount\minute
\newtoks\amorpm
\hour=\time\divide\hour by60
\minute=\time{\multiply\hour by60 \global\advance\minute by-\hour}
\edef\standardtime{{\ifnum\hour<12 \global\amorpm={am}%
        \else\global\amorpm={pm}\advance\hour by-12 \fi
        \ifnum\hour=0 \hour=12 \fi
        \number\hour:\ifnum\minute<10 0\fi\number\minute\the\amorpm}}
\edef\militarytime{\number\hour:\ifnum\minute<10 0\fi\number\minute}

\def\draftlabel#1{{\@bsphack\if@filesw {\let\thepage\relax
      \xdef\@gtempa{\write\@auxout{\string
          \newlabel{#1}{{\@currentlabel}{\thepage}}}}}\@gtempa \if@nobreak
    \ifvmode\nobreak\fi\fi\fi\@esphack} \gdef\@eqnlabel{#1}} \def\@eqnlabel{}
\def\@vacuum{}
\def\draftmarginnote#1{\marginpar{\raggedright\scriptsize\tt#1}}

\def\draft{
  \oddsidemargin -.5truein
  \def\@oddfoot{\footnotesize \sl preliminary draft \hfil
    \rm\thepage\hfil\sl\today\quad\militarytime}
  \let\@evenfoot\@oddfoot \overfullrule 3pt
    \let\label=\draftlabel
    \let\marginnote=\draftmarginnote
  \def\@eqnnum{(\theequation)\rlap{\kern\marginparsep\tt\@eqnlabel}%
    \global\let\@eqnlabel\@vacuum}

  }

\makeatletter
\newdimen\normalarrayskip              
\newdimen\minarrayskip                 
\normalarrayskip\baselineskip
\minarrayskip\jot
\newif\ifold             \oldtrue            \def\new{\oldfalse}
\def\arraymode{\ifold\relax\else\displaystyle\fi} 
\def\eqnumphantom{\phantom{(\theequation)}}     
\def\@arrayskip{\ifold\baselineskip\z@\lineskip\z@
     \else
     \baselineskip\minarrayskip\lineskip2\minarrayskip\fi}
\def\@arrayclassz{\ifcase \@lastchclass \@acolampacol \or
\@ampacol \or \or \or \@addamp \or
   \@acolampacol \or \@firstampfalse \@acol \fi
\edef\@preamble{\@preamble
  \ifcase \@chnum
     \hfil$\relax\arraymode\@sharp$\hfil
     \or $\relax\arraymode\@sharp$\hfil
     \or \hfil$\relax\arraymode\@sharp$\fi}}
\def\@array[#1]#2{\setbox\@arstrutbox=\hbox{\vrule
     height\arraystretch \ht\strutbox
     depth\arraystretch \dp\strutbox
     width\z@}\@mkpream{#2}\edef\@preamble{\halign
\noexpand\@halignto
\bgroup \tabskip\z@ \@arstrut \@preamble \tabskip\z@ \cr}%
\let\@startpbox\@@startpbox \let\@endpbox\@@endpbox
  \if #1t\vtop \else \if#1b\vbox \else \vcenter \fi\fi
  \bgroup \let\par\relax
  \let\@sharp##\let\protect\relax
  \@arrayskip\@preamble}
\def\eqnarray{\stepcounter{equation}%
              \let\@currentlabel=\theequation
              \global\@eqnswtrue
              \global\@eqcnt\z@
              \tabskip\@centering
              \let\\=\@eqncr
 \halign to \displaywidth\bgroup
    \eqnumphantom\@eqnsel\hskip\@centering
    $\displaystyle \tabskip\z@ {##}$%
    \global\@eqcnt\@ne \hskip 2\arraycolsep
         $\displaystyle\arraymode{##}$\hfil
    \global\@eqcnt\tw@ \hskip 2\arraycolsep
         $\displaystyle\tabskip\z@{##}$\hfil
         \tabskip\@centering
    &{##}\tabskip\z@\cr}
\begingroup\ifx\undefined\newsymbol \else\def\input#1 {\endgroup}\fi
\newfont{\hr}{msbm10}
\newfont{\ams}{msam10}

\voffset= - 1.3in
\hoffset= - 1.0in         

\def\beq{\begin{equation}}
\def\eeq{\end{equation}}
\def\ba{\beq\new\begin{array}{c}}
\def\ea{\end{array}\eeq}
\def\be{\ba}
\def\ee{\ea}

\def\Bf#1{\mbox{\boldmath $#1$}}
\def\res{{\rm res}}
\def\F{{\cal F}}
\def\d{\partial}
\def\N2{${\cal N}=2$}
\def\2{{1\over 2}}

\begin{document}
\begin{flushright}
FIAN/TD-12/01\\
ITEP/TH-43/01
\end{flushright}
\vspace{0.5cm}

\begin{center}
\renewcommand{\thefootnote}{${\!}^\star$}
{\LARGE \bf Associativity Equations in Effective
SUSY Quantum Field Theories
\footnote{
Contribution to the proceedings of the conference {\em SUSY01},
Dubna, June 2001 } }
\end{center}
\vspace{0.5cm}

\setcounter{footnote}{0}
\begin{center}
{\large A.~Marshakov}\\
\bigskip
{\em Theory Department, Lebedev Physics Institute, and\\
Institute of Theoretical and Experimental Physics, Moscow, Russia}\\
\medskip
{\sf e-mail:\ mars@lpi.ru, andrei@heron.itep.ru}\\
\bigskip\bigskip\medskip
\end{center}

\begin{quotation}
\noindent
The role of associativity or WDVV equations in effective supersymmetric
quantum theories is discussed and it is demonstrated that for
wide class of their solutions when residue formulas are valid the
proof of associativity equations can be reduced
to solving the system of ordinary linear equations and depends only upon
corresponding matching and nondegeneracy conditions. The covariance of
WDVV equations upon generic duality transformations and the role of
associativity equations in general context of quasiclassical integrable
systems is also discussed.
\end{quotation}

\section{Introduction: effective nonperturbative SUSY theories}

More than 30 years have passed after pioneering work of Golfand and Likhtman
\cite{GL}
but supersymmetry (SUSY) is still among both most hot and intriguing topics of
modern theoretical physics. Looking around one may distinguish two main
directions of investigations of supersymmetry, the first is devoted to
the search of the (signs of) SUSY in real nature and the second mostly
concerns supersymmetry as a very effective theoretical laboratory for study
mainly nonperturbative effects in quantum field theories and string theory.
While the first direction is still supported only by beleif that
superpartners will be found sometimes on colliders, the second direction
is based on more solid ground, since on pure theoretical level SUSY has
become already a part of reality in physics. Indeed,
for example, it is enough to
require world-line (or world-sheet in string theory) supersymmetry
in the first-quantized theory
to get the space-time fermions from the space-time bosons.

The main topic of this talk is to discuss some strong (though very indirect)
outcomes of
SUSY for the effective nonperturbative quantum field and string
theories. Our starting point is that, speaking
in sigma-model terms, SUSY determines or strongly restricts the
{\em geometry} of corresponding effective action, which (dependently
on the number of SUSY generators) necessarily possesses complex
(K\"ahler, special K\"ahler, hyper-K\"ahler etc) structure. This is pure
kinematic fact and it is very important since it is preserved if one goes
beyond perturbation theory where conventional quantum field theory does
not work. Beyond weak coupling one may rely only upon some
heuristic tools, like duality, which should be, of course, consistent with
the properties of corresponding target-space complex geometry.

Hence, in what follows we almost
forget about fermions and Lagrangians of SUSY gauge theories and
will not distinguish between SUSY and corresponding geometry it induces on moduli space.
A possible way to formulate any nonperturbative statements comes from
the relation between complex geometry and theory of integrable systems --
nonlinear integrable differential equations. We will discuss one of the most
intriguing examples of these equations which arises in the context of SUSY
effective theories -- the associativity or WDVV equations
\cite{WDVV} and demonstrate in particular that they are indeed consistent with
properties of corresponding geometry underlying SUSY gauge theories.

For moduli spaces of vector multiplets in \N2 SUSY gauge theories this
geometry is (a rigid analog of) special K\"ahler \cite{speKa,SW}.
It means that metric on moduli space is determined by a single
{\em holomorphic} function $\F ({\bf a})$ of several
complex arguments (VEV of complex matrix scalar field). This function can
be determined even non-perturbatively \cite{SW}, i.e. beyond the scope of
standard quantum gauge theory. It means
that there should exist some other way to define nonperturbative effective
theory and one of the possibilities to do this is to use the above relation
between SUSY and geometry and connections between geometry of complex
manifolds and theory of nonlinear integrable differential equations.

From this point of view it is very important that function
$\F$ satisfies some well-known nonlinear integrable differential equations
\cite{GKMMM} and, in particular, the set of
associativity or
WDVV equations which can be written in the form \cite{MMM}
\be
\label{WDVV}
\F_i\cdot \F_j^{-1}\cdot\F_k = \F_k\cdot \F_j^{-1}\cdot\F_i
\qquad \forall i,j,k \;,
\ee
for the matrices $\|\F_i\|_{jk} \equiv \F_{ijk}$ whose
matrix elements are the third derivatives of the function $\F({\bf a})$,
\be
\label{thirdder}
\F_{ijk} = {\d^3\F\over\d a_i\,\d a_j\,\d a_k}
\ee
Originallly WDVV equations were found \cite{WDVV} in the context of
two-dimensional topological (\N2 twisted superconformal) theory. However,
it was shown later \cite{MMM} that the WDVV equations
are much more universal and can be extended
at least to the \N2 SUSY gauge theories in four dimensions. Despite
simplicity of their compact matrix form (\ref{WDVV}) they form indeed an
overdetermined system of highly nontrivial
nonlinear differential equations satisfied by function
$\F$, apart of the case of functions of one or two variables, when equations
(\ref{WDVV}) are empty, i.e. are satisfied by any function.

\section{WDVV equations: associative algebra and residue formulas}

In two-dimensional topological theories, for example
\N2 SUSY Landau-Ginzburg models, which are defined by
superpotential $W$ so that vacua are identified with $dW=0$,
the WDVV equations arose \cite{WDVV} as consequence of the crossing relations
\be
\label{crossing}
\sum_k C^k_{ij}C^n_{kl} = \sum_k C^k_{il}C^n_{kj}
\ee
for the structure constants of the operator algebra of
primary ot vacuum operators in topological (say, Landau-Ginzburg) model
\be
\label{alg}
\phi_i\cdot\phi_j =\sum_k  C^k_{ij}\phi_k
\ee
Equations (\ref{crossing}) are
algebraic relations and they turn into the system of nonlinear differential
WDVV equations only upon
identification of three-point functions of the operators
$\{ \phi_i \}$ with the third derivatives of some function $\F (a_1,\dots,
a_n)$
\be
\label{3point}
\langle \phi_i\phi_j\phi_k \rangle = {\d^3\F\over\d a_i\,\d a_j\,\d a_k}
\ee
In the Landau-Ginzburg models algebra (\ref{alg}) is realized as ring of
all polynomials modulo $dW=0$ and
formula (\ref{3point}) acquires the form of the {\em residue formula}
\cite{WDVV,KriW,Dub}
\be
\label{res}
{\cal F}_{ijk} = \sum_\alpha \res_{\lambda_\alpha}{\phi_i(\lambda )
\phi_j(\lambda)
\phi_k(\lambda)\over W'(\lambda)Q'(\lambda)} =
\sum_\alpha {\phi_i(\lambda_\alpha)\phi_j(\lambda_\alpha)
\phi_k(\lambda_\alpha)
\over W''(\lambda_\alpha)Q'(\lambda_\alpha)} \equiv
\sum_\alpha \res_{\lambda_\alpha}{dH_idH_jdH_k\over
dWdQ}
\ee
For simplicity we have considered the case when superpotential is function of
a single variable and denoted $\phi_i(\lambda)={dH_i\over d\lambda}$,
$W'(\lambda)={dW\over d\lambda}$
and $Q'(\lambda)={dQ\over d\lambda}$. However, the properties
of the formula (\ref{res})
we are going to use below do not really depend on the number of the
variables, and universality of formula (\ref{res}) goes far beyond the
Landau-Ginzburg case.

It is necessary to stress that {\em both} algebra (\ref{alg}) and residue
formula (\ref{res}) are necessary for the validity of the WDVV equations
(\ref{WDVV}). In sect.~\ref{ss:finite} we demonstrate that imposing
these two conditions is actually {\em enough} to prove (\ref{WDVV}) and almost
nothing extra (except for nondegeneracy) should be added. In various
models algebra (\ref{alg}) can be realized in different ways, but it can be
{\em always} presented as algebra of functions (of course, not necessarily
polynomials) modulo vanishing on some
submanifold. If this submanifold as realized as a set of zeroes of some
differential $dW=0$ this algebra is finite and leads together with residue
formula (\ref{res}) (certainly with the same $dW$ in denominator) to the
finite system of
WDVV equations on function $\F$ of finite number of variables.
We also demonstrate in sect.~\ref{ss:finite} that when residue formla
(\ref{res}) is valid the proof of validity of the WDVV equations is reduced to
the problem of solving the system of ordinary linear equations and for that
all extra ingredients, common in the context of two-dimensional topological
theories \cite{Dub} (constancy of "metric", existence of unity operator etc)
are absolutely inessential.

Finally in this section, let us make few remarks concerning appearence of
the WDVV equations in general context of quasiclassical integrable hierarchies,
i.e. beyond SUSY topological and Seiberg-Witten theories.
It turns out \cite{BMRWZ} that at least in some cases they can be considered
as a direct consequence of (infinite) dispersionless Hirota relations. On one
hand from the point of view of associativity algebras (\ref{alg}) this is a
rather trivial statement since for infinite number of operators it is
much more easy to construct a closed algebra. Nevertheless (see
\cite{BMRWZ} for details) it is possible to show, say in the known
Landau-Ginzburg
case, that the finite WDVV equations (\ref{WDVV}) are consequence of these
infinite ones and, moreover, some new solutions of the associativity
equations can be obtained in this way.

\section{WDVV as solving system of linear equations}
\label{ss:finite}

Require now the differential $dW(\lambda)$ in (\ref{res}) to be
{\em meromorphic} with finite number of zeroes at some points
$\{\lambda_\alpha\}$
\be
W'(\lambda_\alpha)=0
\ee
and $Q'(\lambda_\alpha)\neq 0$.
The only extra condition one should impose below is
that matrix $\| \phi_i(\lambda_\beta)\| $ is non-degenerate
\be
\label{det}
\det_{i\alpha}\| \phi_{i}(\lambda_\alpha)\| \neq 0
\ee
In particular, (\ref{det}) requires "matching" $\#(i)=\#(\alpha)$, i.e. the
number of ``hamiltonians'' $\{ dH_i\}$ or ``fields'' $\{ \phi_i\}$ should be
{\em exactly} equal
to the number of zeroes  $\{ \lambda_\alpha \}$.
One may define now the structure constants $C_{ij}^k$ of this
finite-dimensional algebra (\ref{alg}) (where the sum is finite) from the
system of {\em linear equations}
\be
\label{eqc}
\phi_i(\lambda_\alpha)\phi_j(\lambda_\alpha) =\sum_k
C^k_{ij}\phi_k(\lambda_\alpha), \ \ \ \ \ \ \ \forall\ \lambda_\alpha
\ee
which hold for {\em all} zeroes $\{ \lambda_\alpha\}$ of $dW$.
Formula (\ref{eqc}) gives a realization of the
finite-dimensional associative algebra (\ref{alg}) defined by any
meromorphic differential $dW$. Using matching and nondegeneracy
conditions (\ref{det}), one can simply solve the system (\ref{eqc}) and write
\be
\label{litc}
C^k_{ij} = \sum_\alpha
\phi_i(\lambda_\alpha)\phi_j(\lambda_\alpha)
\left(\phi_k(\lambda_\alpha)\right)^{-1}
\ee
where the last factor means matrix inverse to $\| \phi_i(\lambda_\alpha)\| $.

The situation does not change at all, if instead of (\ref{eqc}) we
consider an {\em isomorphic} algebra
\footnote{The situation here is very similar to considered in
\cite{MMM,MMMlong} in the context of algebra of 1-differentials
on Riemann surfaces. However, in
contrast to algebra of forms, algebra of functions (\ref{eqcxi})
is {\em always} associative.}
\be
\label{eqcxi}
\phi_i(\lambda_\alpha)\phi_j(\lambda_\alpha) =\sum_k
C^k_{ij}(\xi)\phi_k(\lambda_\alpha)\cdot\xi(\lambda_\alpha),
\ \ \ \ \ \ \ \forall\ \lambda_\alpha
\ee
with the only requirement $\xi(\lambda_\alpha)\neq 0$, for $\forall\alpha$,
and (\ref{eqc}) is a particular case of (\ref{eqcxi}) with
$\xi(\lambda)\equiv 1$. Then, instead of (\ref{litc}), one immediately gets
\be
\label{litcxi}
C^k_{ij}(\xi) = \sum_\alpha
{\phi_i(\lambda_\alpha)\phi_j(\lambda_\alpha)\over\xi(\lambda_\alpha)}
\left(\phi_k(\lambda_\alpha)\right)^{-1}
\ee
When the algebra (\ref{eqcxi}) leads to finite WDVV {\em equations}
(\ref{WDVV})?
In order to get the answer one should check consistency between formulas
(\ref{litcxi}) and (\ref{res}), which has the form
\be
\label{feta}
{\cal F}_{ijk} = \sum_l C_{ij}^l(\xi)\eta_{kl}(\xi)
\ee
and ``metric'' $\eta_{kl}(\xi)$ (which
depends upon $\xi $ in order to cancel dependence of the structure constants)
is non degenerate and satisfies
\be
\label{metric}
\eta_{kl}(\xi) = \sum_a \xi_a {\cal F}_{kla}
\ee
where the third derivatives ${\cal F}_{kla}$ (\ref{thirdder})
are given by residue formula
(\ref{res}) and $\{\xi_a\}$ are some coefficients, which can even depend
on times. Now, substituting (\ref{res}) into (\ref{metric}) one gets
\be
\eta_{kl}(\xi) =
\sum_\alpha \res_{\lambda_\alpha}{\phi_k(\lambda)\phi_l(\lambda)\xi(\lambda)
\over
W'(\lambda)Q'(\lambda)} =
\sum_\alpha{\phi_k(\lambda_\alpha)\phi_l(\lambda_\alpha)\xi(\lambda_\alpha)\over
W''(\lambda_\alpha)Q'(\lambda_\alpha)}
\ee
where
\be
\label{xifun}
\xi(\lambda ) = \sum_a\xi_a\phi_a(\lambda )
\ee
Since we already required $\xi(\lambda )$ not to have
zeros in the points $\{\lambda_\alpha\}$, using
condition (\ref{det}) one can always find the corresponding $\xi_a$, solving
again the system of linear equations.

The rest is simple matrix algebra, requiring again {\em only} matching
condition $\#(\alpha)=\#(i)$. Write
\be
\sum_k C_{ij}^k(\xi)\eta_{kl}(\xi) =
\sum_\alpha\sum_k\sum_\beta
{\phi_i(\lambda_\alpha)\phi_j(\lambda_\alpha)\over\xi(\lambda_\alpha)}
\cdot\left(\phi_k(\lambda_\alpha)\right)^{-1}\cdot\phi_k(\lambda_\beta)\cdot
{\phi_l(\lambda_\beta)\xi(\lambda_\beta)\over
W''(\lambda_\beta)Q'(\lambda_\beta)}
\ee
and consider it as a product of four matrices. Two mutually inverse
factors in the middle cancel each other and one finally gets
\be
\sum_k C_{ij}^k(\xi)\eta_{kl}(\xi) = \sum_\alpha
{\phi_i(\lambda_\alpha)\phi_j(\lambda_\alpha)\over\xi(\lambda_\alpha)}
{\phi_l(\lambda_\alpha)\xi(\lambda_\alpha)\over
W''(\lambda_\alpha)Q'(\lambda_\alpha)} =
\sum_\alpha
{\phi_i(\lambda_\alpha)\phi_j(\lambda_\alpha)\phi_l(\lambda_\alpha)\over
W''(\lambda_\alpha)Q'(\lambda_\alpha)} = {\cal F}_{ijl}
\ee
and it means that algebra (\ref{eqcxi}) leads to the WDVV equations
(\ref{WDVV}).
Note that derivation is valid for {\em any} function $\xi(\lambda )$ with
the only
restriction that $\xi(\lambda_\alpha)\neq 0$ and, thus, constancy metric is
absolutely inessential. When all time dependence is
hidden into differential $dW$ the matching condition is satisfied
automatically, at least if $W(\lambda)$ is a polynomial
(the Landau-Ginzburg case). Below we also consider two other examples
when, in contrast to the Landau-Ginzburg case, the matching condition is
at least naively violated into one or another direction. In the first case
(e.g. Seiberg-Witten prepotential for softly broken ${\cal N}=4$ Yang-Mills
theory) one should necessarily add extra variables to the Seiberg-Witten
periods, in the second case (one of the examples is given by tau-functions
of conformal maps \cite{BMRWZ}) the situation is even more striking: $\F =
\log\tau$ satisfies the WDVV equations as a function of only {\em part} of
its variables, when the rest of the variables is fixed.

\section{Consistency in the examples of Seiberg-Witten prepotentials
and tau-functions of curves}

In the Seiberg-Witten theory (the pure \N2 Yang-Mills or Toda chain case; on
correspondence between Seiberg-Witten theories and integrable systems see
\cite{GKMMM}) the residue formula has the form \cite{MMM}
\be
\label{ressw}
{\cal F}_{ijk} =
\oint_{{dw\over w}=0} {d\omega_id\omega_jd\omega_k\over d\lambda{dw\over w}}
= \sum_\alpha \res_{\lambda_\alpha}{d\omega_id\omega_jd\omega_k\over
d\lambda {P_N'(\lambda)d\lambda\over y}} = \\ =
\sum_\alpha
\res_{\lambda_\alpha}{\phi_i(\lambda)\phi_j(\lambda)\phi_k(\lambda)
\over y^2P_N'(\lambda)} =
\sum_\alpha \res_{\lambda_\alpha}{\phi_i(\lambda)\phi_j(\lambda)
\phi_k(\lambda)\over (P_N(\lambda)^2-4)P_N'(\lambda)}
\ee
where the Seiberg-Witten curve is \cite{sun}
\be
\label{c1}
w + {1\over w} = P_N(\lambda)
\ee
or
\be
\label{c2}
y^2 = P_N(\lambda)^2 - 4
\ee
with $P_N(\lambda)$ being a polynomial of degree $N$ and $y = w - {1\over
w}$. In (\ref{ressw}) the role of hamiltonians $\{ dH_i\}$ from (\ref{res})
is played by the set of canonical holomorphic differentials on Riemann
surface (\ref{c1}), (\ref{c2})
\be
\label{holdif}
d\omega_i = {\phi_i(\lambda)d\lambda\over y}
\ee
where $\{ \phi_i(\lambda)\}$ are certain polynomials of degree not exceeding
$(N-2)$. The set of these polynomials is supposed to be nondegenerate in
the points $\{\lambda_\alpha\}$,
which are the zeroes of ${dw\over w} = {dP_N(\lambda)\over y}$,
i.e. solutions to $P_N'(\lambda_\alpha)=0$.

We know, that despite the difference between formulas (\ref{res}) and
(\ref{ressw}) the WDVV equations (\ref{WDVV}) do hold for the Seiberg-Witten
theory \cite{MMM}. The reason is the same as in sect.~\ref{ss:finite}:
the matching condition between
the number of holomorphic differentials $d\omega_i$ and the zeroes of
$P_N'(\lambda)$ holds {\em exactly} since both numbers are
equal to the genus of Riemann surface (\ref{c1}), (\ref{c2}) which is $g=N-1$.
And this is {\em all} (together with (\ref{det})) we need for derivation
of (\ref{WDVV}) from (\ref{ressw}), the proof literally repeats that of
sect.~\ref{ss:finite}.

Now, why this may cause difficulties in more general situation in
Seiberg-Witten theory \cite{MMMlong}? The reason is that, by derivation,
formula (\ref{ressw}) is {\em accidental}, since what was really
derived in \cite{MMM} is
\be
\label{ressw1}
{\cal F}_{ijk} = -
\oint_{d\lambda=0} {d\omega_id\omega_jd\omega_k\over d\lambda{dw\over
w}} =
\oint_{{dw\over w}=0} {d\omega_id\omega_jd\omega_k\over d\lambda{dw\over w}}
\ee
where the second integral (used in (\ref{ressw}) to be rewritten in the
form of (\ref{res})) is a consequence of the first due to holomorphic
properties of the integrand. The first equality in (\ref{ressw1}) was
obtained by differentiating the period matrix $T_{ij}=\F_{ij}$ of the
Riemann surface (\ref{c1}), (\ref{c2}) and using some relations for the
values of canonical holomorphic differentials (\ref{holdif}) at the branch
points of hyperelliptic curve, defined by $d\lambda=0$ (and {\em
not} by ${dw\over w}=0$). However, the number of hyperelliptic branch
points $d\lambda=0$, as follows from (\ref{c2}), equals to $2N$ and
the matching condition naively would {\em fail}! What saves the situation
is that, using that integrand in (\ref{ressw1}) is holomorphic, one can
rewrite the same contour integral around the zeroes
${dw\over w} = {dP_N(\lambda)\over y} =0$, which
is still not enough, since the number of zeroes of ${dw\over w}$ is
$2g=2(N-1)$, but due to hyperelliptic ${\bf Z}_2$-symmetry of exchanging
$\lambda$-sheets one can finally bring the residue formula to the form of
(\ref{ressw}), i.e. as a sum over $g=N-1$ zeroes of
the polynomial $P_N'(\lambda)$ (each of them corresponds in fact to the
pair of points on curve (\ref{c1}), (\ref{c2})). It means that the matching
condition for the Seiberg-Witten Toda chain case finally holds!

Now, it becomes clear that for other prepotentials on nontrivial Riemann
surfaces the matching conditions may fail \cite{MMMlong}.
For example, in the case of elliptic Calogero-Moser or
broken ${\cal N}=4$ Seiberg-Witten theory the generating
differential, instead of $\lambda {dw\over w}$
is $\lambda dz$, where $dz$ is canonical holomorphic differential on base
torus and function $\lambda$ satisfies the Lax equation
\be
\label{CaMo}
\det_{N\times N}(\lambda - L(z)) = 0
\ee
with the Lax operator $L(z)$ introduced in \cite{KriCal}. The number of
zeroes of $d\lambda$ and $dz$ can be calculated from the Riemann-Roch
theorem, saying, in particular, that for any meromorphic differential
\be
\# ({\rm zeroes}) - \#({\rm poles}) = 2g-2 = 2N-2
\ee
since the genus of the curve
(\ref{CaMo}) is $g=N$. The differential $dz$ is holomorphic
(it is holomorphic on base torus and does not acquire poles on the cover),
so one gets
\be
\label{zedz}
\# ({\rm zeroes}\ dz) = 2N-2
\ee
quite similar to its analog ${dw\over w}$ in Toda case (\ref{c1}), (\ref{c2}).
However, we
do not have anymore the hyperelliptic symmetry, which allows to "reduce factor
$2$" and, say, rewrite
(\ref{ressw}) as a sum over only $(N-1)$ points ({\em half} of
$(2N-2)$). As for the second differential $d\lambda$, since it follows from
(\ref{CaMo}) and the properties of the elliptic Calogero-Moser Lax
operator \cite{KriCal} that
\be
d\lambda \sim {dz\over z^2}
\ee
it has $N$ second-order poles, hence
\be
\# ({\rm zeroes}\ d\lambda) - \#({\rm poles}\ d\lambda) =
\# ({\rm zeroes}\ d\lambda) - 2N = 2N-2
\ee
or
\be
\# ({\rm zeroes}\ d\lambda) = 4N-2
\ee
i.e. it has even more zeroes than (\ref{zedz}). It means that restricting
(\ref{res}) to the case of holomorphic differentials only
\be
\label{ressw2}
{\cal F}_{ijk} = -
\oint_{d\lambda=0} {d\omega_id\omega_jd\omega_k\over d\lambda dz} =
\oint_{dz=0} {d\omega_id\omega_jd\omega_k\over d\lambda dz}
\ee
one would get
$\#(\alpha)>\#(i)$ and in order to
close the algebra, corresponding to (\ref{ressw2})
one needs to add to the set of $N$ holomorphic
differentials at least $(N-2)$ extra ``hamiltonians''. Naively there are
two direct options to do that -- to add either meromorphic differentials with
higher order poles or non single-valued holomorphic differentials in spirit
of \cite{KriW}. One needs then, however, to check the (extended) residue
formula (\ref{ressw2}) with added new meromorphic or non single-valued
differentials.
From the point of view of SUSY quantum theory the main problem is the
physical sense of corresponding extra time variables which should play the
role of "hidden" moduli parameters in corresponding Seiberg-Witten theory.

The opposite example (which has no yet known direct relation to SUSY apart
of being presented as "generalized" topological theory) is given by
dispersionless tau-functions, in particular
by tau-functions of analytic curves or conformal maps \cite{CM}. These
tau-functions
satisfy well-known algebraic Hirota
relations for the second derivatives
\be
\label{second}
\F_{ij} \equiv {\d^2 \F\over\d a_i\d a_j}
\ee
(dispersionless limit of the bilinear Hirota relations) and
(infinite) WDVV equations can be derived taking times-derivatives of algebraic
Hirota relations \cite{BMRWZ}. The residue formula (\ref{res}) in this case
can be also thought of as a consequence of Hirota relations.

One may consider tau-functions corresponding to rational
conformal maps \cite{CM} (tau-function of curves restricted to the finite
number of time variables). It is easy to see that in this case matching
condition will be violated into the opposite (compare to Seiberg-Witten
examples) direction, so that $\#(\alpha)<\#(i)$. It means that (logarithm of)
tau-function of rational conformal map satisfies the WDVV equation
(\ref{WDVV}) as a function of {\em any} (in the case of corresponding
non-degeneracy) {\em part} of its variable whose total number is equal to
$\#(\alpha)$ or to the number of zeroes of differential of rational map.
Unfortunately in the only explicit case of ellipse
there cannot be any reasonable WDVV since algebra (\ref{alg}) is
two-dimensional so that equations (\ref{WDVV}) are empty, but if one manages
to find explicit formulas for the other tau-functions of rational maps,
their logarithms would give examples of functions, satisfying equations
(\ref{WDVV}) as functions of {\em part} of their variables.

\section{Conclusions and outlook}

We have tried to demonstrate in this talk that when conventional quantum
field theory fails to describe nonperturbative effective theories one can
still use the relation between SUSY and geometry and even encode the
corresponding
geometric information in the solutions to some very specific nonlinear
integrable differential equations. As an example of such equations we have
considered the WDVV equations (\ref{WDVV}) and demonstrate that their
existence is based on two very universal facts -- associative algebra and
residue formula. The corresponding proof, presented above, is extremely
simple and does not at all involve any additional requirements like
constancy of "topological metric", existence of unity operator,
(see \cite{Dub}) etc.
However, there is still no {\em direct} link between the
WDVV equations and first principles of non-perturbative physics and finally
we would like to make few comments on the possible connection.

\begin{itemize}

\item{\bf Duality}

From the point of view of relation between SUSY and geometry it is very
important that
WDVV equations (\ref{WDVV}) are consistent with generic duality
transformations \cite{dWM}
\be
\F^S ({\bf a}^S) =  \F({\bf a})  + {\textstyle {1\over2}} {\bf
a}^t\cdot U^t \cdot W\cdot {\bf a} + {\bf a}^t\cdot W^t \cdot Z \cdot
\left({\d\F\over\d{\bf a}}\right)  + {\textstyle {1\over2}}
\left({\d\F\over\d{\bf a}}\right)^t\cdot Z^t \cdot V\cdot
\left({\d\F\over\d{\bf a}}\right)  \,. \label{fsdu}
\ee
with matrices $U$, $V$, $W$ and $Z$ forming symplectic group or obeying
relations
\be
\label{uvwz}
U^t\cdot V-Z^t\cdot W = V\cdot U^t-Z\cdot W^t = 1\,,
\\
U^t\cdot W = W^t\cdot U\,,\qquad Z^t\cdot V = V^t\cdot Z\,.
\ee
The proof of this fact is very simple and
requires only trivial matrix algebra. However, the outcomes are very
important: the WDVV equations are internally consistent with
properties of special K\"ahler geometry and their form (\ref{WDVV}),
where only third derivatives of $\F$, but not (special K\"ahler) metric
appear, is covariant
under transformations which should be necessarily respected by
nonperturbative physics. It is also very interesting to understand the sense
of transformations (\ref{fsdu}) in general context of quasiclassical
integrable hierarchies and (generalized) dispersionless Hirota equations.

\item{\bf From WDVV to algebraic relations on second derivatives}

In the case of dispersionless hierarchies, the {\em infinite} WDVV equations
follow from the dispersionless Hirota relations \cite{BMRWZ} for the second
derivatives (\ref{second}), which can be written in the form
\be
\label{hir}
\F_{ij} = T_{ij}({\Bf\varphi})
\ee
where $T_{ij}$ are some known functions and $\{\varphi_i\}$ denote some
restricted set of the second derivatives, say
\be
\label{vf}
\varphi_i = \F_{1i}({\bf a})
\ee
The structure constants (\ref{alg}) for the WDVV equations can be defined as
\be
\label{c}
C_{ij}^k = {\d T_{ij}\over\d\varphi_k}
\ee
Indeed, then
\be
\F_{ijk} = {\d \F_{ij}\over\d a_k} = \sum_l {\d T_{ij}\over\d\varphi_l}
{\d\varphi_l\over\d a_k} = \sum_l C_{ij}^l\F_{1kl}
\ee
which repeats (\ref{feta}) with some special "topological metric"
\cite{WDVV,Dub}
$\eta_{ij} = \F_{1ij}$. In dispersionless hierarchies crossing relations
(\ref{crossing}) for the structure constants (\ref{c}) follow from bilinear
Hirota identities (\ref{hir}), but naively this is true only for infinite
number of variables ${\bf a} = (a_1,a_2,\dots)$ and equations.

In the essentially finite dimensional case (say, prepotentials \N2 SUSY
gauge theories in
four dimensions) the second derivatives of $\F$ (\ref{second})
form symmetric $N\times N$ matrix with $\2 N(N+1)$ generally different
elements $\F_{ij}({\bf a})$ and
in fact we have $\2 N(N+1)-N = \2 N(N-1)$ nontrivial functions $T_{ij}
({\Bf\varphi})$.
The WDVV equations (\ref{crossing}), (\ref{WDVV})
become now the set of first-order differential equations for these
functions
\be
\label{eqt}
\sum_l {\d T_{ij}\over\d\varphi_l} {\d T_{lk}\over\d\varphi_n} =
\sum_l {\d T_{ik}\over\d\varphi_l} {\d T_{lj}\over\d\varphi_n}
\ee
The question we are going to address is whether
(\ref{eqt}) may be rewritten in some sensible form -- for example as
{\em algebraic} relations on $T_{ij}$ similar to dispersionless
Hirota equations. It is not clear even
in the simplest nontrivial $N=3$ case ($N=1$ and $N=2$ are trivial as they
should be). If such algebraic relations exist they may play the role of RG
relations for the theory with many couplings and/or some analogs of the
Schottkey relations for the matrix elements of the (Seiberg-Witten) period
matrices.

\end{itemize}

\section*{Acknowledgements}
I am grateful to A.Boyarsky, H.Braden, B.~de~Wit, L.Hoevenaars,
I.Krichever, A.Mironov, A.Morozov, S.Na\-tan\-zon, O.Ru\-chay\-skiy,
P.Wiegmann and A.Zabrodin
for many important discussions and to D.Kazakov and other organizers
of {\em SUSY01} for warm hospitality in Dubna. The work was
partially supported by RFBR grant
No.~01-01-00539, INTAS grant No.~97-0103, CRDF grant No.  RP1-2102
(6531) and the grant for the support of scientific schools
No.~00-15-96566.

\end{document}

\end{document}